\documentclass[11pt]{article}

\usepackage{euscript}
\usepackage{amssymb}
\usepackage{amsfonts}
\usepackage{amsbsy}
\usepackage{amsmath}
\usepackage{epsfig}
\usepackage{amsthm}
\usepackage{amscd}
\usepackage{amstext}
\usepackage{verbatim}
\usepackage[numbers,sort&compress]{natbib}

\usepackage[usenames,dvipsnames,table]{xcolor}
\usepackage[pdftex]{hyperref}

\def\ben{\begin{equation}}
\def\een{\end{equation}}
\def\half{{\textstyle{1\over2}}}

    \let\e=\varepsilon
   
  \let\n=\nu  \let\p=\phi

    \let\L=\Lambda
 \let\P=\Phi

\def\be{\begin{equation}}
\def\ee{\end{equation}}
\def\beq{\begin{equation}}
\def\eeq{\end{equation}}
\def\ba{\begin{array}}
\def\ea{\end{array}}

\def\e{\epsilon}

\def\dalemb#1#2{{\vbox{\hrule height .#2pt
       \hbox{\vrule width.#2pt height#1pt \kern#1pt
               \vrule width.#2pt}
       \hrule height.#2pt}}}
\def\square{\mathord{\dalemb{6.8}{7}\hbox{\hskip1pt}}}
\newcommand{\bea}{\begin{eqnarray}}
\newcommand{\eea}{\end{eqnarray}}
\newcommand{\tr}{{\rm tr} }

\def\Lag{{\mathcal{L}}}

\def\EE{{\mathcal{E}}}

\renewcommand{\p}{\partial}

\newcommand{\refe}[1]{(\ref{#1})}

\def\FTheta{{\bf{\Theta}}}

\def\Qxi{{\bf{Q}}_\xi}
\def\FTheta{{\bf{\Theta}}}
\def\L0vac{{\cal L}_0^{vac}}
\def\P0vac{{\cal P}_0^{vac}}
\def\ffL0{{\cal L}_0}
\def\ffP0{{\cal P}_0}
\def\SWald{S_{\text{Wald}}}
\def\SWCFT{S_{\text{WCFT}}}
\def\fL{{\bf L}}
\def\fS{{\bf S}}
\def\fJ{{\bf J}}
\def\fM{{\bf M}}
\def\fQ{{\bf Q}}
\def\fH{{\bf H}}

\def\fQ{{\bf{Q}}}

\begin{document}

\begin{center}

{ \LARGE {\bf Warped AdS$_3$ Black Holes in Higher Derivative Gravity Theories}}

\vspace{1.2cm}

St\'{e}phane Detournay$^\sharp$, Laure-Anne Douxchamps$^{\sharp,\flat}$, Gim Seng Ng$^\flat$ and C\'eline Zwikel$^\sharp$ 

\vspace{0.9cm}

\centerline{\sl $^\sharp$Universit\'e Libre de Bruxelles and 
International Solvay Institutes}
\centerline{\sl $^\sharp$Physique Th\'eorique et Math\'ematique, }
\centerline{\sl Campus Plaine C.P. 231, B-1050 Bruxelles, Belgium}
\smallskip

\vspace{0.5cm}

{\it $^\flat$ Department of Physics, McGill University, Montr\'eal, QC H3A 2T8, Canada \\}

%

\vspace{1.6cm}

{\tt sdetourn@ulb.ac.be, ldouxcha@ulb.ac.be, gim.ng@mcgill.ca, czwikel@ulb.ac.be} \\

\vspace{1.6cm}

\end{center}

\begin{abstract}
We consider warped AdS$_3$ black holes in generic higher derivatives gravity theories in 2+1 dimensions. The asymptotic symmetry group of the phase space containing these black holes is the semi-direct product of a centrally extended Virasoro algebra and an affine $u(1)$ Kac-Moody algebra. 
Previous works have shown that in some specific theories, the entropy of these black holes agrees with a Cardy-like entropy formula
derived for warped conformal field theories.  In this paper, we show that this entropy matching continues to hold for the most general higher derivative theories of gravity. We also discuss the existence of phase transitions.

\end{abstract}

\pagebreak
\setcounter{page}{1}

\tableofcontents

\pagebreak

\section{Introduction and Summary}
Two-dimensional conformal field theories (2d CFTs) and their celebrated Cardy formula play a central role in our attempt at understanding the microscopic degrees of freedom responsible for black hole entropy. Strominger and Vafa's seminal derivation \cite{Strominger:1996sh} of the entropy of the three-charges extremal D1-D5 black hole revealed that the low energy effective field theory governing the excitations of the system flows in the infrared to a 2d CFT, whose asymptotic number of states counted by the Cardy formula exactly accounts for the Bekenstein-Hawking entropy. This result was eventually generalized to a variety of other setups, including both extremal and near-extremal higher-dimensional black holes \cite{Callan:1996dv,Peet:1998wn, Cvetic:1999ja,Cvetic:1996gq,Horowitz:1996ay,Maldacena:1996gb,Johnson:1996ga,Maldacena:1997de}, and generic BTZ black holes \cite{Banados:1992wn,Banados:1992gq} in three dimensions \cite{Strominger:1997eq}. The matching between black hole and Cardy entropies could be traced back to the presence of an AdS$_3$ factor (in the near-horizon region for the higher-dimensional cases) of the corresponding geometries. The result followed as a consequence of the work of Brown and Henneaux \cite{Brown:1986nw} who showed that the classical phase space of AdS$_3$ gravity is endowed with the action of the 2d local conformal group, implying that states of the corresponding quantum gravity theory should organize into representations of a 2d CFT \cite{Strominger:1997eq}.

More precisely, the Cardy formula is a well-known universal property of unitary and modular invariant 2d CFTs encapsulating the asymptotic degeneracy of states at large charges/temperatures for fixed central charges:
\bea \label{Cardy}
  S_{Cardy} &=& 2 \pi \sqrt{\frac{c^+ L_0^+}{6}} + 2 \pi \sqrt{\frac{c^- L_0^-}{6}} \\
     &=& \frac{\pi^2}{3} (c_+ T^+ + c_- T^-).
\eea
Remarkably, this formula matches the Bekenstein-Hawking entropy of the BTZ black holes \emph{for all values of the charges/potentials} (and not only $L_0^\pm \rightarrow \infty$ or $T^\pm \rightarrow \infty$). The reason for this seemingly unreasonable effectiveness was clarified in \cite{Hartman:2014oaa} (see also \cite{Hellerman:2009bu, Hellerman:2010qd, Keller:2011xi} for related earlier works): for CFTs with a large central charge (corresponding to the semi-classical limit on the gravity side) and a sparse light spectrum, the Cardy regime extends all the way down to where the condition $L_0^+ + L_0^- \geq \frac{c}{12}$ is satisfied.
This behavior puts constraints on the precise nature of 2d CFTs potentially dual to AdS$_3$ theories of gravity, examples of which include symmetric orbifolds \cite{David:2002wn, Pakman:2009zz, Avery:2010er, Avery:2010hs, Keller:2011xi, Hartman:2014oaa} and extremal CFTs \cite{Witten:2007kt,Maloney:2007ud}. Other CFTs potentially dual to AdS$_3$ gravity have recently been discussed in \cite{Belin:2015hwa}
The key property responsible for the existence of a simple and elegant formula like (\ref{Cardy}) is modular invariance. For a partition function at inverse temperature $\beta$, it is expressed by the equality
\be \label{ModInv}
  Z(\beta) = Z\left(\frac{4 \pi^2}{\beta}\right).
\ee
It is the latter expression that allows to express the high temperature partition function using only minimal data about  the vacuum of the theory. The universal character of the asymptotic growth of the number of states in 2d CFTs, as captured by the Cardy formula, is a consequence of the peculiar UV/IR connection in 2d CFTs embodied in \refe{ModInv}, relating high/low-temperature regimes of the theory. A similar behaviour is expected to explain the universality of the area-entropy law in gravity theories. On the one hand, the Bekenstein-Hawking entropy $ S_{BH} =$ Area$/4G_N$ captures to leading order in $1/G_N$ the UV density of states of a quantum gravity theory governed by the Einstein action in the IR. On the other hand, it is a reasonable expectation that, beyond its mere existence, the specific details of a UV-complete gravity theory are actually not needed to understand the area law, much like the actual quantum theory of atoms was not needed to derive the universal laws of thermodynamics for gases from statistical reasoning \cite{Bredberg:2011hp}.

In any sensible approach to quantum gravity, the Einstein-Hilbert action will only provide the leading low-energy effective action.
At high energies, it is expected to be corrected by higher-curvature terms suppressed by a dimensionful scale. Typically in string theories, the coupling would correspond to the inverse string tension $\alpha'$. In that case, the entropy-area law is no longer expected to hold, and is modified in order for the first law of black hole mechanics to hold 
\cite{Wald:1993nt,Iyer:1994ys,Iyer:1995kg}. Interestingly, the match between Cardy and black hole entropies still persists \cite{Kraus:2005vz, Behrndt:1998eq, LopesCardoso:2000fp, LopesCardoso:2000qm, LopesCardoso:1999za, LopesCardoso:1999xn, LopesCardoso:1999fsj, LopesCardoso:1999cv, LopesCardoso:1998tkj, Maldacena:1997de, Harvey:1998bx,  Saida:1999ec,Kraus:2006wn}. The effect of higher-curvature corrections in AdS$_3$ gravity boils down to a global multiplicative renormalization of all the charges of the theory.



Of course, AdS$_3$ spaces and asymptotically AdS boundary conditions are believed to be part of just one (although possibly the simplest) among many manifestations of the holographic principle, and intensive work has been devoted to depart from the AdS/CFT paradigm. This includes for instance dS/CFT \cite{Spradlin:2001pw}, AdS/Condensed Matter Theory \cite{Hartnoll:2009sz, Hartnoll:2011fn}, 
Non-relativistic Schr\"odinger and Lifshitz Holography \cite{Taylor:2015glc}, Hyperscaling geometries \cite{Huijse:2011ef}, Flat Space Holography \cite{Afshar:2014rwa} and Kerr/CFT \cite{Bredberg:2011hp, Compere:2012jk} (references are by no means meant to be extensive, and point most of the time at reviews on the subjects).

In this work, we will focus on one particular example of non-AdS Holography: Warped AdS$_3$ spaces (WAdS$_3$) and Warped Conformal field Theories (WCFTs) \cite{Israel:2004vv, Moussa:2003fc, Anninos:2008fx,Detournay:2012pc,Hofman:2011zj}. The former have appeared in various contexts, in relation to Kerr/CFT and Schr\"odinger holography, as UV deformations of AdS$_3$ spaces. They are parameterized by two constants which are usually taken as $\ell$, the original AdS$_3$ radius, and $(\nu^2 - 1)$ characterizing the departure from AdS$_3$. They were observed to display an Asymptotic Symmetry Group (ASG) different from the Brown-Henneaux conformal symmetry group, consisting instead in the semi-direct product of a Virasoro algebra and a $u(1)$ affine Kac-Moody algebra \cite{Compere:2007in, Compere:2008cv, Compere:2009zj,Blagojevic:2009ek, Henneaux:2011hv}. WCFTs are defined as two-dimensional field theories with precisely these symmetries. Interestingly, these theories possess an infinite-dimensional symmetry group as well as a notion of modular invariance allowing the derivation of a Cardy-type formula for the density of states given by \cite{Detournay:2012pc} 
\bea \label{swcftintro} 
\SWCFT &=&-4\pi i\frac{P_0 P_0^{vac}}{k}+2\pi\sqrt{\frac{c}{6}\left(L_0
-\frac{P_0^2}{k}\right)} \\
&=& \frac{2\pi i  }{\Omega} P_0^{vac}-\frac{8\pi^2}{\beta\Omega} \left(\frac{(P_0^{vac})^2}{k} - \frac{c}{24}\right) \label{swcftv2intro} 
\eea 
where $L_0$ and $P_0$ are the zero-modes of the Virasoro and Kac-Moody (KM) current generators respectively, $\beta$ the inverse temperature and $\Omega$ the angular velocity, $c$ is the Virasoro central charge and $k$ the $u(1)$ level. $P_0^{vac}$ is the charge of the vacuum state on the cylinder which is not fixed by the symmetries of the theory through a plane-to-cylinder map like in 2d CFTs\footnote{Eq. (\ref{Cardy}) also implicitly contains the vacuum charges on the cylinder, but in a unitary CFT these are given by $L_0^{\pm, vac} = -\frac{c^\pm}{24}$. In WCFTs, only $L_0^{vac}$ is constrained by unitarity bounds to be $L_0^{vac} = \frac{(P_0^{vac})^2}{k} - \frac{c}{24}$.}. Further properties of these new classes of field theories have been explored e.g. in \cite{Castro:2015uaa, Castro:2015csg}.  Remarkably, the horizon entropy of the black hole solutions belonging to the WAdS$_3$ phase space was seen to exactly match the field theory expression (\ref{swcftintro}). Note that similar counting formulas and matchings have been observed in Lifshitz \cite{Gonzalez:2011nz, Bravo-Gaete:2015iwa}, Flat Space \cite{Barnich:2012xq, Bagchi:2012xr}, Hyperscaling \cite{Shaghoulian:2015dwa} and Rindler \cite{Shaghoulian:2015dwa, Afshar:2015wjm} holography in 2+1 space-time dimensions, but also in higher dimensions \cite{Shaghoulian:2015lcn}.

So far, the matching between Bekenstein-Hawking and field theory entropies for WAdS$_3$ black holes has been performed on a case-by-case basis in specific theories admitting WAdS$_3$ spaces as a solution -- these backgrounds do not solve the vacuum Einstein's equation, so they require including matter fields or higher curvature corrections (see e.g. \cite{Detournay:2012pc} for Topologically Massive Gravity \cite{Deser:1981wh, Deser:1982vy} and string theory embeddings, \cite{Donnay:2015iia, Ghodsi:2010gk} for New Massive Gravity \cite{Bergshoeff:2009hq}, \cite{Ghodsi:2010ev, Ghodsi:2011ua} for Born-Infeld extensions of NMG). 
The goal of the present work will be to extend this to a completely arbitrary higher-curvature gravity theory admitting WAdS$_3$ as a solution. We will indeed show that the black hole entropy is always reproduced by (\ref{swcftintro}) or (\ref{swcftv2intro}).
The result is comparable in spirit to the analogous statement for BTZ black holes that their entropy is always reproduced by a Cardy formula \cite{Saida:1999ec, Kraus:2005vz}, and similarly in Kerr/CFT where the original derivation of \cite{Guica:2008mu}
could be extended to an arbitrary higher-curvature theory \cite{Azeyanagi:2009wf}. We will closely follow the philosophy of the latter work. On the one hand, we will compute the black hole entropy of the WAdS$_3$ black holes given a general gravity theory with a Lagrangian $L$. It is known that in the presence of higher-derivative corrections, the entropy is no longer given by the horizon area, but rather by Wald formula \cite{Wald:1993nt, Iyer:1995kg, Jacobson:1993vj, Iyer:1994ys} which in 3d reads as
\be\label{eq:Swald}
\SWald=-2 \pi\int_0^{2\pi}
Z^{\alpha\beta\mu\nu}\epsilon_{\alpha\beta} \epsilon_{\mu\nu} \sqrt{g_{\phi\phi}} |_{r=r_+} \,,
\ee  where  $Z^{\alpha\beta\mu\nu}$ is constructed out of curvature invariants of the metric and possesses the symmetries of the Riemann tensor and $\epsilon_{\alpha\beta}$ is the binormal at the horizon $r_+$. In the case where the Lagrangian is only a function of the Riemann (without covariant derivatives), then symbolically $Z\sim \partial L/\partial Riemann$. In general, the expression is more complicated (see Sec.~\ref{Sect3} for details). 

Applying Eq.~(\ref{eq:Swald}) to the WAdS$_3$ black holes,  since all these black holes have a local $SL(2,\mathbb{R}) \times U(1)$ isometry, this will allow us to rewrite the Wald entropy in a very simple form as
\be\label{eq:Swaldintro}
\SWald= -\frac{64  \pi ^2 \nu ^2}{\left(-3+5 \nu ^2\right)  }\frac A\Omega
\ee 
where 
$A$ is a constant depending only on $(\nu,\ell)$ and the couplings of the theory (in particular, they do not depend on the black hole parameters). Then, exploiting again the symmetries and using the Covariant Phase Formalism \cite{Iyer:1994ys, Azeyanagi:2009wf, Lee:1990nz, Wald:1993nt, Barnich:2001jy,Barnich:2007bf, Compere:2007az} we will be able to compute the charges and central charges appearing in (\ref{swcftintro}) and (\ref{swcftv2intro}). In particular, we will find that
\begin{align}
& P_0^{vac}=\frac{32\, i \,\pi \, \nu ^2}{-3+5 \nu ^2}\,A,\\
& k=-\frac{32  \,\pi \, \nu \, \left(3+\nu ^2\right)}{\ell \left(-3+5 \nu ^2\right)}\,A,\\
& c=\frac{768  \, \ell \,\pi \, \nu ^3}{\left(3+\nu ^2\right) \left(-3+5 \nu ^2\right)}\,A
\end{align}
which, when plugged in (\ref{swcftv2intro}), exactly reproduces (\ref{eq:Swaldintro}):
\be
\SWald=\SWCFT\,.
\ee 
In the body of the paper we will further discuss equality in other ensembles.

As a by-product of our computations, which allowed us to compute the charges of the Warped Black Holes in arbitrary theories, we reconsider the existence of a Hawking-Page phase transition for these backgrounds, generalizing the results of \cite{Detournay:2015ysa} in Topologically Massive Gravity.

The paper is organized as follows. 
We will first review in Sect.~\ref{eq:WadsWCFT} the phase space and asymptotic symmetry group of asymptotically WAdS$_3$ spacetimes as well as the WCFT entropy formula in WAdS/WCFT holography.  The warped black hole entropy in New Massive Gravity will be provided to review some known results of the entropy matching.  Moving on, in Sect.~\ref{eq:symmetryWads3}, we further investigate the geometry of WAdS$_3$ and argue that all curvature invariants can be expressed in a manner that will allow us to derive, in Sect~\ref{SectWald}, the Wald entropy of the warped black hole in a very concise form (see Eq.~(\ref{eq:Swaldintro})). Moreover, these simplifications will make possible the computation of the charges for an arbitrary theory in Sect.~\ref{sec:charges}. 
Sect.~\ref{sec:examples} is devoted to provide explicit formulae in a few examples in a hope to make the arguments more understandable and transparent. In the final section, we discuss the Hawking-Page phase transition for WAdS$_3$ black holes. Technical details are relegated to two appendices.

\section{Review: WAdS$_3$ Spaces WCFT Entropy Matching in NMG }
\label{eq:WadsWCFT}
%

In this section, we review some basic features about WAdS$_3$ spaces and reproduce the match between Bekenstein-Hawking and WCFT entropies in the case of New Massive Gravity.


\subsection{Phase Space}

WAdS$_3$ spaces \cite{Rooman:1998xf, Israel:2004vv, Anninos:2008fx} can be obtained by deforming the AdS$_3$ metric in the Kerr-Schild-like manner,
breaking the $SL(2,\mathbb{R})\times SL(2,\mathbb{R})$ isometry group of AdS$_3$ down to a $SL(2,\mathbb{R})\times U(1)$ isometry group. 
The metric in global coordinates is
\be ds^2=\frac{\ell^2}{\nu^2+3}\left[-\cosh^2(\sigma)d\tau^2+d\sigma^2+\frac{4\nu^2}{\nu^2+3}\Big(du+\sinh(\sigma)d\tau\Big)^2\right]\ee
in which we recognize the AdS$_3$ metric (for $\nu = 1$) but otherwise with a squashing/stretching factor $4\nu^2/(\nu^2+3)$.
In this case, black hole solutions exist and are quotients of the WAdS$_3$ metric. The black hole metric can be written in the so-called warped-black-hole coordinates of \cite{Compere:2009zj} as:
\be \label{eq:wbhmetric}
ds^2=dt^2+\frac{dr^2}{\frac{r^2}{\ell^2}(\n^2+3)-12mr+\frac{4j\ell}{\nu}}+d\phi^2\left(\frac{3r^2}{\ell^2}(\n^2-1)+12m r-\frac{4j\ell}{\nu}\right)+dt\,d\phi\left(\frac{-4\nu r}{\ell}\right) \ee
with $r\in [0,\infty)$, $t\in (-\infty,\infty)$, $\phi\sim \phi+2\pi$, and $m$ and $j$ are parameters characterizing the black hole. That metric is a black hole with two horizons in the case where $j<\frac{9 \ell m^2 \nu }{3+\nu ^2}$. 
They are located at
\be r_{\pm}=\frac{2\ell^2}{\nu^2+3} \left( 3m\pm \sqrt{9m^2-\frac{j}{\nu\ell}(\nu^2+3)}\right).\ee 
When the warp parameter $\nu^2>1$, the solution is said to be stretched, and when $\nu^2<1$, it is squashed. For $\nu^2=1$ we recover locally AdS$_3$ space. We will focus on the spacelike stretched case, which exhibits no pathologies such as naked closed timelike curves, unlike its squashed counterpart. 

An important feature of these metrics is that they are part of exact string theory backgrounds, like AdS$_3$ spaces. The latter are known to be the target space of an exact Conformal Field Theory, an $SL(2,\mathbb{R})$ WZW model \cite{Horowitz:1993jc,Balog:1988jb,Petropoulos:1989fc,Bars:1990rb,Hwang:1990aq,Maldacena:2000hw}. These models provide exact solutions to all orders in $\alpha'$ modulo a renormalization of the WZW level, identified with the AdS radius \cite{Tseytlin:1992ri}. WAdS$_3$ spaces, on the other hand, are part of backgrounds representing marginal deformations of the $SL(2,\mathbb{R})$ WZW model \cite{Detournay:2005fz}. It was shown in Sect 2.3 of \cite{Detournay:2010rh} that the background fields extracted from the classical action get renormalized only through a redefinition of the parameters in the metric (the AdS radius and squashing parameter). Therefore, when considering WAdS$_3$ black holes in higher-curvature theories, we will have to consider the modification to their entropy only due to replacing the area law by Wald's entropy (and not to a modification of the geometry).


\subsection{Asymptotic Symmetries}

WAdS$_3$ spaces are not asymptotically AdS$_3$ (except of course for $\nu^2=1$) and do not belong to the Brown-Henneaux phase space \cite{Brown:1986nw}. Instead, they satisfy the following boundary conditions \cite{Compere:2009zj,Compere:2007in,Compere:2008cv, Blagojevic:2009ek, Henneaux:2011hv} (coordinates are $(t,r,\phi)$):
\be g_{BC}=\begin{pmatrix}
1+O(r^{-1})&O(r^{-2})&\frac{-2\nu r}{\ell}+O(r^0)\\
O(r^{-2})&\frac{\ell^2}{\nu^2+3}\frac{1}{r^2}+O(r^{-3})&O(r^{-1})\\
\frac{-2\nu r}{\ell}+O(r^0)&O(r^{-1})&\frac{3(\nu^2-1)}{\ell^2}r^2+O(r)\\
\end{pmatrix}
\ee

%
%


The infinitesimal diffeomorphisms leaving these boundary conditions invariant are  generated by the asymptotic Killing vectors \cite{Compere:2009zj} 
\bea \nonumber
l_n &=&e^{in\phi}\left(\left(1 +O(r^{-1})\right)\partial_t +\left(-inr +O(1)\right)\partial_r+ \left(1+O(r^{-2})\right)\partial_{\phi}\right)\\ \label{eq:asympKV} 
p_n &=&e^{in\phi}\left(1+O(r^{-1})\right)\partial_t. \eea
These generators obey the following Lie-commutation relations:
\be i[l_m,l_n]=(m-n)\,l_{m+n}, \qquad i[l_m,p_n]=-n\,p_{m+n}, \qquad [p_m,p_n]=0\,.\ee
The conserved charges $L_m$, $P_m$ associated to these generators $l_n,p_n$ satisfy a Virasoro-Kac-Moody $U(1)$ algebra:
\begin{align}
& i \{L_m,L_n\}=(m-n)L_{m+n}+\frac c{12}(m^3-m)\delta_{m+n,0}\\
& i \{L_m,P_n\}=-nP_{m+n}\\
& i \{P_m,P_n\}=\frac k2  \,m\delta_{m+n,0}\, .
\end{align} The explicit definitions and expressions of charges (in particular the central charge and the level) and their commutators will be provided later, since they are (bulk)-theory-dependent.

\subsection{WCFT and BH Entropies in NMG}

The above algebra is the defining symmetry algebra of a WCFT, a two-dimensional field theory with global $SL(2,\mathbb{R})\times U(1)$ invariance \cite{Hofman:2011zj, Detournay:2012pc}. These theories share many features with 2d CFTs, in particular the existence of a formula counting the degeneracy of states at large potentials/charges, given in the micro-canonical ensemble by \cite{Detournay:2012pc}



\be \label{swcft} S_{\textrm{WCFT}}=-4\pi i\frac{P_0P_0^{vac}}{k}+2\pi\sqrt{\frac{c}{6}\left(L_0-\frac{P_0^2}{k}\right)}\,.
\ee
This formula has allowed to match entropies on the gravity and field theory sides in various theories. We briefly review here the case of NMG \cite{Donnay:2015iia}.

The NMG action is given by
\begin{equation}
I_{NMG}=\frac1{16\pi} \int d^3x\sqrt{-g}\left(   (R-2\Lambda) +\frac1{p} \left( R_{\mu\nu}R^{\mu\nu} -\frac38 R^2 \right) \right)
\end{equation}
where $p$ has the dimensions of a mass square. Warped black holes are solutions of NMG for the following couplings:
\begin{equation}
p= \frac{3-20 \nu ^2}{2 \ell^2}, \qquad \Lambda = \frac{4 \nu ^4-48 \nu ^2+9}{2 \ell^2 \left(20 \nu ^2-3\right)}.
\end{equation}
In terms of the angular velocity
\begin{equation}
\Omega=-\frac{\nu ^2+3}{4 \left(\sqrt{\ell \,\nu  \left(9 \ell m^2 \nu -\left(\nu ^2+3\right) j
     \right)}+3 \ell m \nu \right)}
\end{equation}
the entropy can be expressed as
\begin{equation}
S_{\textrm{BH}}=-\frac{8 \pi  \nu ^2}{\left(20 \nu ^2-3\right) \Omega }
\end{equation}
on the gravity side. On the field theory side, the central charge $c$ and level $k$ in the asymptotic symmetries algebra are
\begin{equation}\label{central-charges-NMG}
c=\frac{96\ell\nu^3}{(3+\nu^2)(20\nu^2-3)}, \qquad k=-\frac{4\nu(3+\nu^2)}{\ell(20\nu^2-3)}.
\end{equation}
The WCFT formula for the entropy can be rewritten in terms of $\Omega$ and the inverse temperature $\beta$ as
\begin{equation}
S_{\textrm{WCFT}}=\frac{2\pi i  }{\Omega} P_0^{vac}-\frac{8\pi^2}{\beta\Omega}L_0^{vac}
\end{equation}
with 
\begin{equation}
L_0^{vac}=\frac{(P_0^{vac})^2}{k}-\frac {c}{24} .
\end{equation}
$P_0$ is the charge associated to $\partial_t$ and $L_0$ the one associated to $\partial_{\phi}$. The vacuum is obtained by taking $m= i /6$ and $j=0$, corresponding to the only metric in the family (\ref{eq:wbhmetric}) with enhanced $SL(2,\mathbb{R})\times U(1)$ isometry and no conical defect in appropriate coordinates (for a detailed discussion, see \cite{Detournay:2012pc}), which  yields 
\begin{equation}
P_0^{vac}=\frac{4 \nu ^2 i}{20 \nu ^2-3}, \qquad L_0^{vac}=0. 
\end{equation}
We then have
\begin{equation}
S_{\textrm{BH}}=S_{\textrm{WCFT}}
\end{equation}
as expected. 

The goal of this paper is to show that this matching holds for any higher-derivative gravity theory.

\section{Geometry of WAdS$_3$ and Wald Entropy}\label{Sect3}

\subsection{Symmetries and Curvature Tensors}
\label{eq:symmetryWads3}
In this section, we summarize some key facts related to curvature invariants constructed out of a WAdS$_3$ metric. In particular, these results are applicable to the warped black holes since they are quotients of the WAdS$_3$ metric.

First of all, in 3d, all Riemann tensors can be expressed in terms of Ricci tensors.
In the case of WAdS$_3$, from \cite{Anninos:2008fx}, we know that some products of Ricci tensors are
\be
\left\{
R,R_{\mu\nu} R^{\mu\nu},
R_{\mu\nu}R^{\nu\rho}R_{\rho}{}^{\mu}
\right\}
=\frac{6}{\ell^2}\left\{
-1,\frac{\nu^4-2\nu^2-3}{\ell^2},
\frac{-\nu^6-3\nu^4+9\nu^2-9}{\ell^4}
\right\}\,.
\ee This can be easily verified for the metric Eq.~(\ref{eq:wbhmetric}).
Furthermore, recall that in a maximally symmetric spacetime, all curvature tensors (for e.g. product of covariant derivatives of Riemann/Ricci tensors) are expressible (covariantly) in terms of the metric tensor.\footnote{For example, $R_{\mu\nu\rho\sigma}=\text{const} \left(
g_{\mu\rho} g_{\nu\sigma}
-g_{\nu\rho} g_{\mu\sigma}
\right)$, from which all products of (covariant derivatives) of Riemanns can hence be constructed in terms of the metric tensor.} In the present case where we only have $SL(2,\mathbb{R})\times U(1)$, a slight generalization of such an expression is available as well. This is in essence the argument in  Appendix D of \cite{Azeyanagi:2009wf}.

Any  tensor constructed out of the metric should respect the $SL(2,\mathbb{R})\times U(1)$ isometry. The consequences have been investigated and exploited in Ref.~\cite{Azeyanagi:2009wf}.
In particular, any {\it scalar} curvature invariants constructed out of the metric are constants. As such, in our case, these constants can only depend on $\nu$ and $\ell$ and not on the parameters of the black holes (i.e. $m$ and $j$), which are parameters of the global quotients\footnote{There are two ways to see this. First, repeating the argument in Appendix D.1 of Ref.~\cite{Azeyanagi:2009wf}, one writes down a Killing vector which takes a particular point in the manifold to any other point. Then one uses a Lorentz boost in the tangent space to argue that all curvature scalars are invariants under this boost. These two facts imply that the value of any curvature scalar is constant over the manifold. Alternatively, a more direct way to see this is to evaluate the Ricci tensor and its covariant derivatives for the black hole spacetimes (i.e. Eq.~(\ref{RdR})). In three dimensions, all curvature scalars are just products of covariant derivatives of Ricci with appropriate contractions. Using Eq.~(\ref{dJ})-(\ref{RdR}), upon contracting all indices, one sees that eventually one ends up with just constants depending on $\nu$ and $\ell$.}.

For a general tensor, let us consider what happens in the case of a symmetric-two tensor $S^{\mu\nu}$. Due to boost-invariance (which is a consequence of the $SL(2,\mathbb{R})\times U(1)$ symmetry), in a conveniently chosen vielbein basis $e^{\hat{0}},e^{\hat{1}}$ and $e^{\hat{2}}$ (we wrote these as one-forms $e^{\hat{i}}\equiv e^{\hat{i}}{}_\mu dx^\mu$), which is given explicitly in  \cite{Azeyanagi:2009wf},
we have \be
S^{\hat{0}\hat{0}}=-S^{\hat{1}\hat{1}},\quad S^{\hat{0}\hat{1}}=S^{\hat{0}\hat{2}}=S^{\hat{1}\hat{2}}=0
\ee while $S^{\hat{2}\hat{2}}$ is arbitrary. This implies that any such tensor only contains two arbitrary components.
In particular, we can conveniently decompose it as
\be
S^{\hat{a}\hat{b}} = c_1 \eta^{\hat{a}\hat{b}}+c_2 J^{\hat{a}} J^{\hat{b}}\,.
\ee where the constants $c_i$'s only depend on $\nu$ and $\ell$. The vector $J^\mu$ is most usefully chosen to be the $U(1)_R$ of the $SL(2,\mathbb{R})_L\times U(1)_R$, 
\be
J^\mu\partial_\mu = \partial_t=p_0\, ,
\ee where we work in the warped black hole coordinates in which Eq.~(\ref{eq:wbhmetric}) is written and $p_0$ is the $U(1)_R$ Killing vector (see Eq.~(\ref{eq:asympKV})). Note that $J^\mu J_\mu=1$.
Translating back into spacetime indices, we obtain
\be
S^{\mu\nu} = c_1 g^{\mu\nu} + c_2 J^\mu J^\nu\,.
\ee

Furthermore, note that
\be \label{dJ}
\nabla_\mu J_\nu =\frac{\nu}{\ell} \epsilon_{\mu\nu \sigma} J^\sigma
\ee where the convention is $\epsilon_{tr\phi}=\sqrt{-g} =1$.
Thus, all products of covariant derivatives of $S$ can in turn be rewritten as products of $g$, $\epsilon$ and $J$. 
Let us give an example where $S_{\mu\nu}=R_{\mu\nu}$. Then,
\be \label{RdR}
R_{\mu\nu}=\frac{\nu^2-3}{\ell^2} g_{\mu\nu}
+3\frac{1-\nu^2}{\ell^2} J_\mu J_\nu\, ,\quad
\nabla_\mu R_{\nu\rho}=
3\nu\frac{\nu^2-1}{\ell^3}\left[
\epsilon_{\mu\nu\sigma} J_\rho 
+\epsilon_{\mu\rho\sigma}J_\nu 
\right] J^\sigma\,.
\ee

Throughout this paper, we will be interested in a tensor $Z^{\alpha\beta\mu\nu}$ which has the same index-symmetry as the Riemann tensor and is constructed out of the metric (and its derivatives). It also depends on the theory one considers. By general $SL(2,\mathbb{R})\times U(1)$ symmetric arguments, we can repeat the above arguments to obtain
\be \label{Z}
Z^{\alpha\beta\mu\nu} = A\left[
g^{\mu\alpha}   g^{ \beta \nu} 
-g^{\alpha\nu} g^{\beta \mu}  
\right]+
B\left[
g^{\mu\alpha}   R^{\beta\nu}
-g^{\nu\alpha} R^{\beta \mu}
+g^{\beta\nu} R^{\alpha \mu}
-g^{\beta\mu} R^{\alpha\nu}
\right]
\ee for some constants $A$ and $B$ that only depend on $(\nu,\ell)$. In this case, we prefer to reexpress $J_\mu J_\nu$ in terms of $R_{\mu\nu}$ such that the RHS is expressed in terms of product of $g^{\mu\nu}$ and $R^{\mu\nu}$.

On the other hand, the equations of motion always take the form
\be
K_{\mu\nu}[g_{\alpha\beta}]=0
\ee where $K_{\mu\nu}[g_{\alpha\beta}]$ for a given symmetric-two-tensor constructed out of the metric. For e.g. in pure Einstein theory, it is just the Einstein tensor or the Ricci tensor.  Following the above logic, evaluating $K_{\mu\nu}$ on a warped AdS or black hole solution, the symmetry of the geometry allows us to decompose $K_{\mu\nu}$ into a sum of the metric and the Ricci tensor:\footnote{In the case where the Lagrangian contains only Ricci tensors (and not covariant derivatives of Ricci tensor), we work this out very explicitly in Appendix~\ref{app:on-shell}.  For e.g., see Eq.~(\ref{eq:eom11}).}
\be
K_{\mu\nu}=
E_1 R_{\mu\nu}
+E_2 g_{\mu\nu}=0.
\ee where $E_1$ and $E_2$ are constants which only depend on $(\nu,\ell)$ and on the couplings $\alpha_i$ of the theories.
Note that the dependence on the couplings of the theories is linear.
 The equations of motion then reduce to {\it two} independent equations, setting
\be
E_1 (\nu, \ell, \alpha_i) = E_2 (\nu, \ell, \alpha_i) = 0.
\ee This means that as long as we have a theory with two independent couplings, such as that in TMG or NMG, then we will always be able to solve the equations of motion.
The couplings appear linearly in the $E_i$'s, for that reason we can always solve these two decoupled equations (subject to obtaining real $\ell$ and $\nu$ as solutions).

\subsection{Wald Entropy of Warped Black Holes}\label{SectWald}
For any diffeomorphism covariant theory of gravity, the Wald entropy formula in 3d is 
\be
\SWald=-2 \pi\int_0^{2\pi}
\left.  d\phi \, Z^{\alpha\beta\mu\nu}\epsilon_{\alpha\beta} \epsilon_{\mu\nu} \sqrt{g_{\phi\phi}} \right|_{r=r_+}
\ee  
where 
\begin{equation}\label{Zabcdgeneraldef}
Z^{\alpha\beta\mu\nu}=\frac{\delta^{cov}L}{\delta R_{\alpha\beta\mu\nu}} =: \sum_{i=0}(-1)^i\nabla_{(e_1}\cdots \nabla_{e_i)} \frac{\partial L}{\partial\nabla_{(e_1}\cdots \nabla_{e_i)} R_{\alpha\beta\mu\nu}}\,
\end{equation}
and $\epsilon_{\mu\nu}$ is the binormal at the horizon, given by $\epsilon_{\mu\nu} = \nabla_\mu \xi_\nu$ with $\xi$ the generator of the horizon with its surface gravity normalized to unity \cite{Brustein:2011gu}.

In general, the expression of the $Z$ tensor is rather complicated.
However, we have seen that by general $SL(2,\mathbb{R})\times U(1)$ symmetry arguments (see Sec.~\ref{eq:symmetryWads3}), it can be written as \eqref{Z}. 
 Given $Z$ of the form above, we can compute
\bea
Z^{\alpha\beta\mu\nu}\epsilon_{\alpha\beta} \epsilon_{\mu\nu} \sqrt{g_{\phi\phi}} |_{r=r_+}
&=&-4\left[
A
+
B R^\alpha{}_\beta n^\beta{}_\alpha\right]\sqrt{g_{\phi\phi}}|_{r=r_+}
\eea where we have used $\epsilon^{\mu\nu}\epsilon_{\mu\nu}=-2$ and defined $n^\mu{}_\nu\equiv-\epsilon^{\alpha\mu}\epsilon_{\alpha\nu}$.
Furthermore, using
\be
R^\alpha{}_\beta n^\beta{}_\alpha |_{r=r_+} = \frac{2 \left(\nu ^2-3\right)}{\ell^2}
\ee 
 and $\sqrt{g_{\phi\phi}} |_{r=r_+} = -\Omega^{-1}$, one is led to
\bea
\SWald&=&
 \left(32 G_N\pi A\right)\left[
1+\frac{B}{A}\frac{2(\nu^2-3)}{\ell^2}
\right]\times \frac{\text{Area}}{4G_N} 
\eea or
\be \label{Swald}
\SWald= -\frac{ 16\pi^2}{\Omega}
A\left[
1+\frac{B}{A}\frac{2(\nu^2-3)}{\ell^2}
\right]\,.
\ee 
\section{Charges in Warped Black holes}
\label{sec:charges}

\subsection{Covariant Phase Space Formalism}

The definition of (asymptotically) conserved charges is a subtle procedure that can be approached by many differents roads. We will briefly review below the salient features of the formalisms we will be using throughout this work, and refer the reader to the literature for a more detailed summary (see e.g. Sections 3 of \cite{Compere:2009dp} or \cite{Azeyanagi:2009wf}, App. A of \cite{Anninos:2011vd}).

From the Lagrangian $n$-form $\fL$ (with $n$ the spacetime dimension), which is a local functional of all fields denoted collectively $\Phi$, the equations of motion $\EE(\Phi)=0$ are determined as follows:
\be \delta \fL (\Phi) =\EE(\Phi) \delta \Phi+d\FTheta[\delta \Phi,\Phi] \label{eq49}
\ee 
where $\FTheta[\delta \phi,\phi]$, the symplectic potential $(n-1)$-form, can always be chosen to be covariant \cite{Iyer:1994ys}. We use a slight abuse of notation, writing $\EE(\Phi)$ for both an $n$-form and its Hodge dual.
We will be concerned here with the situation where $\Phi$ only consists in the metric tensor $g$.
In that case, the gauge symmetries of the theory are transformations $\delta_\xi g = \Lag_\xi g$ under which the transformation of the Lagrangian is computed as
\be \label{eq50}
  \delta_\xi \fL = \Lag_\xi \fL = d(i_\xi \fL) := d \fM_\xi (\Phi)\,;\quad \fM_{\xi}\equiv i_{\xi} \fL\,. 
\ee
Under such a gauge transformation, one can rewrite, using Bianchi identities,
\be \label{eq51}
 \EE(\Phi) \delta_\xi \Phi = d\fS_\xi(\EE(\Phi),\Phi).
\ee
More explicitly, in a pure gravity theory one has $\EE(g) \delta_\xi g = \nabla_\mu (2 \xi_\nu \EE^{\mu \nu}) - 2 \xi_\nu \nabla_\mu \EE^{\mu \nu}$, where the last term vanishes by virtue of Bianchi identities, and hence in that case\footnote{The one-form $S$ denotes the Hodge dual of the (n-1)-form $\fS$.} $S_\xi^\mu = 2 \xi_\nu \EE^{\mu \nu}$.
$\fS$ enjoys the important property that it vanishes on-shell, and hence is called {\itshape weakly vanishing Noether current}. On the other hand, writing (\ref{eq49}) for a gauge transformation ($\delta = \delta_\xi$) and using (\ref{eq50}), one gets 
\be \label{eq52}
 \EE(\Phi) \delta_\xi \Phi = -d\fJ_\xi(\Phi)
\ee
where the {\itshape canonical Noether current} is defined as
\be \label{eq53}
 \fJ_\xi(\Phi) = \FTheta[\delta_{\xi} \Phi,\Phi] - \fM_\xi (\Phi)
\ee
From (\ref{eq51}) and (\ref{eq52}), one therefore gets that the $(n-1)$-form $\fS_\xi(\EE(\Phi),\Phi) + \fJ_\xi(\Phi)$ is off-shell closed and thus exact, and there exists an $(n-2)$-form $\fQ_\xi (\Phi)$ such that 
\be \label{eq54}
  \fS_\xi(\EE(\Phi),\Phi) + \fJ_\xi(\Phi) \approx \fJ_\xi(\Phi) := -d \fQ_\xi (\Phi)
\ee
where $\approx$ stands for on-shell equality.
$\fQ_\xi (\Phi)$ is the Noether charge as defined by Wald \cite{Wald:1993nt}, which is not to be confused with the conserved charge generating the action of the symmetry generator $\xi$ on the covariant phase space (denoted $\fH_\xi$ or $H_\xi$ for its Hodge dual henceforth). The latter is obtained as follows. Using (\ref{eq50}), (\ref{eq53}),(\ref{eq54}) and the property $\Lag_\xi = i_\xi d + d i_\xi$, the variation of the weakly vanishing Noether current can be expressed \emph{on-shell} as
\be \label{eq55}
\delta \fS_\xi(\EE(\Phi),\Phi) \approx \omega(\delta_\xi \Phi, \delta \Phi) + dk^{IW}_\xi (\delta \Phi, \Phi), 
\ee
where we defined the symplectic structure
\be \label{eq56}
\omega[\delta_1 \Phi,\delta_2 \Phi ;\Phi]=\delta_1\FTheta[\delta_2 \Phi,\Phi]-\delta_2\FTheta[\delta_1 \Phi,\Phi]
\ee
and $(n-2)$-form
\be \label{eq57}
  k^{IW}_\xi (\delta \Phi, \Phi) = -i_\xi \FTheta[\delta \Phi,\Phi] - \delta \fQ_\xi (\Phi)
\ee
Eq. (\ref{eq55}) expresses that when the equations of motion ($\EE(\Phi) = 0$) and the linearized equations of motion ($\delta \EE(\Phi) = 0 = \delta \fS_\xi(\EE(\Phi),\Phi)$) hold, and when $\xi$ is a symmetry ($\delta_\xi \Phi=0$), $ k^{IW}_\xi (\delta \Phi, \Phi)$ defined a conserved charge ($d k^{IW}_\xi (\delta \Phi, \Phi) \approx 0$) given by
\be\label{eq58}
  \delta H_\xi = \int_{C = \partial \Sigma}  k^{IW}_\xi (\delta \Phi, \Phi),
\ee
where $C$ is a Cauchy surface, computing the infinitesimal charge difference between configurations $\Phi$ and $\Phi + \delta \Phi$. Finite charge differences $H_\xi$ are obtained by an integral in configuration space.

An important observation is that the definitions in (\ref{eq55}) are ambiguous up to the redefinitions $$\omega (\delta_\xi \Phi, \delta \Phi) \rightarrow \omega (\delta_\xi \Phi, \delta \Phi) - dE(\delta_\xi \Phi, \delta \Phi),\qquad  k^{IW}_\xi (\delta \Phi, \Phi) \rightarrow  k^{IW}_\xi (\delta \Phi, \Phi) + E(\delta_\xi \Phi, \delta \Phi)$$ for an arbitrary $E(\delta_\xi \Phi, \delta \Phi)$ anti-symmetric in $\delta_\xi \Phi$ and  $\delta \Phi$. This ambiguity generalizes the one in the 
symplectic potential $(n-1)$-form under $\FTheta \rightarrow \FTheta + d{\bf Y}$, and hence in the symplectic structure. One proposal to fix this ambiguity \cite{Barnich:2001jy, Barnich:2007bf} is by acting on the weakly vanishing Noether current with a contracting homotopy operator, yielding an $(n-2)$-form denoted $k^{BB}_\xi (\delta \Phi, \Phi)$.
In essence, this operator is the inverse of the exterior derivative $d$ (see e.g. \cite{Chen:2013aza} for an explicit expression).
One advantage of this procedure is that it provides a definition of charges depending only on the equations of motion
of the Lagrangian, and not on boundary terms. We then have 
\be \label{eq59}
  k^{BB}_\xi (\delta \Phi, \Phi) = k^{IW}_\xi (\delta \Phi, \Phi) + E(\delta_\xi \Phi, \delta \Phi),
\ee
in which the expression of $E(\delta_\xi \Phi, \delta \Phi)$ is known explicitly (see e.g. (3.7) of \cite{Azeyanagi:2009wf}).
Remark that this ambiguity is not relevant for exact symmetries, having $\delta_\xi \Phi = 0$, but may yield distinct results in the asymptotic context (see \cite{Azeyanagi:2009wf} for one such example in Kerr/CFT). In this work, we will be working with Iyer-Wald charges, and explicitly check that the extra term $E$ does not contribute.

Finally, the algebra of charges can be represented by a Dirac bracket as follows:
\be  \label{intkgeneralexpression}
 \delta_\xi H_\zeta := \{H_\zeta, H_\xi\} = H_{[\zeta,\xi]} + \int_{C = \partial \Sigma}  k^{IW}_\zeta (\delta_\xi \Phi, \Phi).
\ee
This is valid on-shell when the charges are integrable. The second term on the right-hand side is recognized as a central extension, which cannot be absorbed in a redefinition of the generators.

\subsection{Expressions for $\Theta$ and $\delta Q_\xi$ }

Following \cite{Azeyanagi:2009wf}, $\FTheta$ and $\Qxi$ admit the decomposition:
\bea\label{expressionsThetaandQ}
\FTheta_{a_2\ldots a_n}&=&\FTheta^{(0)}_{a_2\ldots a_n}+\sum_{s\ge1} \FTheta^{(s)}_{a_2\ldots a_n}\nonumber\\
(\Qxi)_{c_3\ldots c_n}&=&(\Qxi^{(0)})_{c_3\ldots c_n}+\sum_{s\ge 1} (\Qxi^{(s)})_{c_3\ldots c_n}\eea 
where $\FTheta^{(0)}$, $\Qxi^{(0)}$, $\FTheta^{(s)}$ and $\Qxi^{(s)}$ are given by the general expressions \cite{Azeyanagi:2009wf}:
\bea 
\FTheta^{(0)}_{a_2\ldots a_n}&\equiv &-2\left( Z^{abcd} \nabla_d \delta g_{bc} - (\nabla_d Z^{abcd})\delta g_{bc} \right)\epsilon_{a a_2\ldots a_n}\nonumber\\
(\Qxi^{(0)})_{c_3\ldots c_n}&\equiv&\left(- Z^{abcd} \nabla_c\xi_d -2 \xi_c(\nabla_d Z^{abcd}) \right) \epsilon_{abc_3\ldots c_n}. \eea 
\begin{multline}
\FTheta^{(s)}_{a_2\ldots a_n}=\Big[2\left(Z^{ibcd|e_1...e_{s-1}a}+Z^{abcd|e_1...e_{s-1}i}\right)\delta g_{ij}\mathbb{R}^j_{\:\:bcd|e_1...e_{s-1}} 
-2Z^{ibcd|e_1...e_{s-1}j}\delta g_{ij}\mathbb{R}^a_{\:\:bcd|e_1...e_{s-1}}\\+(s-1)\Big(Z^{kbcd|e_1...e_{s-2}ia}\delta g_{ij}\mathbb{R}^{\quad\:\:\:\: j}_{kbcd|\:\:e_1...e_{s-2}}
-\frac{1}{2}Z^{kbcd|e_1...e_{s-2}ij}\delta g_{ij}\mathbb{R}^{\quad\:\:\:\: a}_{kbcd|\:\:e_1...e_{s-2}}\Big)\\
-Z^{kbcd|e_1...e_{s-1}a}\delta\mathbb{R}_{kbcd|e_1...e_{s-1}} \Big] \epsilon_{aa_2\ldots a_n}$$ 
\end{multline}
\begin{multline}(\Qxi^{(s)})_{c_3\ldots c_n}=-2\xi_k\Big[Z^{klcd|e_1...e_{s-1}a}\mathbb{R}^{b}_{\:\:lcd|e_1...e_{s-1}}+Z^{alcd|e_1...e_{s-1}b}\mathbb{R}^{k}_{\:\:lcd|e_1...e_{s-1}}\\
+Z^{alcd|e_1...e_{s-1}k}\mathbb{R}^{b}_{\;lcd|e_1...e_{s-1}}
+\frac{s-1}{2}Z^{lmcd|e_1...e_{s-2}ka}\mathbb{R}^{\quad\quad b}_{lmcd|\:\:e_1...e_{s-2}}\Big]\epsilon_{abc_3\ldots c_n}
\end{multline}
for a $n$-dimensional Lagrangian with $k$-th derivatives of the Riemann tensor ($s=1,\ldots k$). In these expressions, 
the $Z^{abcd}$, $\mathbb{R}_{abcd}$, $Z^{abcd|e_1...e_{s}}$ and $\mathbb{R}_{abcd|e_1...e_{s}}$  are auxiliary fields in terms of which our original Lagrangian can be rewritten without derivatives higher than second order (we will not need their explicit expressions, see however Sect. 4 of \cite{Azeyanagi:2009wf} for details).



Furthermore, the relevant {\bf{E}} term for our computations is \cite{Azeyanagi:2009wf}
\begin{equation}
E_{a_3...a_n}[\mathsterling_{\xi_1}g;\mathsterling_{\xi_2}g]=\frac12\left(-\frac32 Z^{abcd}\mathsterling_{\xi_1}g_c^e\wedge \mathsterling_{\xi_2} g_{ed}+2Z^{acde}\mathsterling_{\xi_1}g_{cd}\wedge \mathsterling_{\xi_2} g^b_e \right)\epsilon_{aba_3...a_n}. 
\end{equation} 
This term will never contribute to the charges. Obviously, it is true for exact Killing vectors. For the central charge, we take $\xi_1=l_n$ and $\xi_2=l_{-n}$ (or $t_n$ ans $t_{-n}$ for the level) and use the expression \eqref{Z} of $Z^{abcd}$. As we will integrate over a $t,r=cst$ surface, we only need to consider the $\phi$ component of $E$. A direct computation shows that it is indeed zero.

%
\subsection{Lagrangians without Derivative of the Ricci}

 \subsubsection{Expressions of charges}
First, we will derive the expressions of the charges for the case of a Lagrangian without derivatives of the Ricci. The exact charges depend only on $\Qxi^{(0)}$ and $\FTheta^{(0)}$ and are given by
\cite{Iyer:1994ys}
\begin{equation}
\delta L_0
\equiv-\int_\infty \delta \fQ_{\partial_\phi},
\end{equation}
\begin{equation}
\delta P_0
\equiv
\int_{\infty} \delta \fQ_{\partial_t} + \int_{\infty} i_{\partial_t}  \FTheta
\end{equation}
where the integral is $(n-2)$-dimensional sphere ($t,r$=constant) at spatial infinity. There is no term with $i_{\partial \phi}\FTheta$ in $\delta L_0$ because $\partial_{\phi}$ is assumed to be tangent to this sphere.

 Using the general form of $Z^{abcd}$ \eqref{Z}, an explicit computation leads to 
\begin{align} 
\nonumber
 L_0 & = 
 \frac{32 \pi  \left(A \ell^2+2 B \left(-3+\nu ^2\right)\right)  }{\ell^2} j  
 +\frac{24 \pi  \nu  \left(-1+\nu ^2\right) \left(A \ell^2+2 B \left(-3+5 \nu ^2\right)\right)}{\ell^5} r^2\,, \\\label{exactcharges}
 P_0 & = \frac{48 \pi \left(A \ell^2+2 B \left(-3+\nu ^2\right)\right)}{\ell^2} m
\end{align}
where $A,B$ are the constants entering in \eqref{Z} and depending on the theory.

For the central terms in \eqref{intkgeneralexpression}, we note that it is sufficient to consider the terms proportional to $n$ and $n^3$ for the computation of the level and the central charge, respectively. We get 
 \begin{align}\nonumber
 k & =\left .2 i \int_\infty k_{p_n}^{IW}[\mathsterling_{p_{-n}}\phi,\bar \phi]\right|_{n}
 =-\frac{32 \pi  \nu }{\ell} \left(A +4 B \frac{\left(-3+2 \nu ^2\right)}{\ell^2}\right),\\ \label{centralcharges} 
 c &=\left .12 i \int_\infty k_{l_n}^{IW}[\mathsterling_{l_{-n}}\phi,\bar \phi]\right|_{n^3}
 =\frac{192 \pi  \nu } { \left(3+\nu ^2\right)} \left(A \ell+2 B \frac{\left (-3+\nu^2\right )}{\ell} \right).
 \end{align}


In whole generality, the charges associated with exact Killing vectors of a metric satisfying the equations of motion of a given theory are finite. Therefore, we see from (\ref{exactcharges}) that a theory admitting WAdS$_3$ as solutions must have its constants $A$ and $B$
 related in the following way:
\begin{equation}\label{relationAB}
B= -\frac{A\, \ell^2}{2 \left(-3+5 \nu ^2\right)}.
\end{equation}
In other words, the coupling constants of the considered theory should satisfy the above relation. As expected it turns out it is equivalent to satisfy the equations of motion (see details in Appendix \ref{app:on-shell}). 

Using \eqref{relationAB}, the charges have the following expression
\begin{align}\label{charges-A}
&  L_0=\frac{128  \,\pi \, \nu ^2  }{-3+5 \nu ^2}A\, j\,,\quad  P_0=\frac{192 \, \pi \, \nu ^2}{-3+5 \nu ^2}A\, m, \\
& k=-\frac{32  \,\pi  \,\nu  \left(3+\nu ^2\right)}{\ell \left(-3+5 \nu ^2\right)}A\,,\quad c=\frac{768\, \ell \, \pi  \,\nu ^3}{\left(3+\nu ^2\right) \left(-3+5 \nu ^2\right)}A
\end{align}
with only one constant depending on the considered theory. 
The Iyer-Wald entropy becomes 
\begin{equation}\label{Iyer-Waldwithoutder}
\SWald=-\frac{64  \pi ^2 \nu ^2}{\left(-3+5 \nu ^2\right)  }\frac A\Omega\,
\end{equation}
which is the result advertised in Eq.~(\ref{eq:Swaldintro}).


Note that the expressions of the charges are proportional to their value in NMG, the simplest theory where warped black holes are solutions: 
\begin{equation}
\{L_0, P_0, k, c\}= \left (\frac{8\pi (-3+20 \nu^2)}{(-3+5\nu^2)}A \right )\, \{L^{NMG}_0, P^{NMG}_0, k^{NMG}, c^{NMG}\}. 
\end{equation}

\subsubsection{Warped Cardy formula}
 In this section, we show that the Iyer-Wald and Warped Cardy entropies match.
 The computation of the latter needs us to evaluate the exact charges for the vacuum solution ($m=i/6$ and $j=0$), which are then given by
 \begin{equation}
 L_0^{vac}=0,\quad 
 P_0^{vac}=\frac{32 i \pi  \nu ^2}{-3+5 \nu ^2}A.
 \end{equation}
We substitute these into the WCFT formula
\be
S_{WCFT} = \frac{2\pi i}{\Omega} P_0^{vac} - \frac{8\pi^2}{\beta \Omega} L_0^{vac},
\ee 
and get precisely the Iyer-Wald entropy \eqref{Iyer-Waldwithoutder}:
\be
S_{WCFT} = \frac{-64  \pi^2  \nu ^2}{(-3+5 \nu ^2)\Omega}A=\SWald \,.
\ee 



\subsection{Lagrangians with Derivatives of the Ricci}
\label{sec:withderivatives}
Now we focus on the general case of a Lagrangian with derivatives of the Ricci.
The first change with respect to the case without derivative is the appearance of new terms in $Z^{abcd}$ as given by \eqref{Zabcdgeneraldef}. However, thanks to the symmetries, it will keep the same form (\ref{Z}) provided we replace $A$ and $B$ in \eqref{Z} by some other $\tilde A$ and $\tilde B$.  In addition, the charges $\Theta^{(s)}$ and $Q^{(s)}$ in \eqref{expressionsThetaandQ} are expected to be corrected as well. 
Nevertheless, we show using the $SL(2, \mathbb{R})\times U(1)$ symmetry (see Appendix \ref{sec:appTheta}) that $P_0$ is not corrected and that $L_0^{(s)},c^{(s)},k^{(s)}$ have the following form:
 \begin{align}\label{charges-correction} \nonumber
L_0^{(s)} & =4\pi\left (\frac{4\, a_3\, \nu ^2+a_2\left(3+\nu ^2\right)}{\ell^2}\right ) r^2
 -\left (48 \,a_2 \,m \right )r
 +\frac{16 \,a_2\, \ell }{\nu }j \\ \nonumber
 c^{(s)} & = \frac{192 \left(3 \,a_2\, \pi +a_2\, \pi  \nu ^2+4\, a_3\,\pi  \nu ^2\right) r^2}{\ell^2}-1152\,a_2\, m \pi  r\\
 k^{(s)} & =-16 \pi \,a_3.
 \end{align}
 with $a_2,a_3$ constants depending on $\nu,\ell$.
Thus the charges are \eqref{exactcharges},\eqref{centralcharges} with $A\rightarrow \tilde A$ and $B\rightarrow\tilde{B}$ plus the corrections \eqref{charges-correction}. 
In order to have finite exact charges, the divergences must cancel. It implies
\begin{equation}
a_2=0,\qquad a_3=-\frac{3  \left(-1+\nu ^2\right) \left(\tilde A \ell^2+2 \tilde B \left(-3+5 \nu ^2\right)\right)}{2\nu\,\ell^3}. 
\end{equation}
The purpose of the first equation is to cancel the $r-$divergence. To get rid of the $r^2-$divergence, the second term in \eqref{exactcharges} for $L_0$ should cancel against the first term in \eqref{charges-correction}. 
Imposing that condition, we get the same expression for the exact charges we had in the case without derivative. In consequence, the Warped Cardy formula again matches with the Iyer-Wald entropy. However, finiteness of the Virasoro central charge requires $a_3 = 0$. This is a necessary condition in order to have a well-defined phase space with Vir + u(1) Kac-Moody symmetries. We will take this as a supplementary condition on the coupling constants of the theory (though it is not excluded that this condition could derive from requiring the metric to satisfy the equations of motion).
In this case, all the corrections vanish. Relation \eqref{relationAB} then holds between $\tilde A$ and $\tilde B$ and we get the same charges \eqref{charges-A} as in the case without derivative of the Ricci.

%
%




In conclusion, we have shown that the Iyer-Wald entropy is always reproduced by a Warped Cardy formula using the $SL(2,\mathbb R)\times U(1)$ symmetries and on-shell conditions. We have also found a general explicit expression for the charges.

%
\section{WAdS$_3$ Entropy in Higher Curvature Theories: Examples}
\label{sec:examples}
In this section, we present a few examples with detailed computations. 
\subsection{$L=a R-\Lambda+ b R^2 + c R_{\mu\nu}R^{\mu\nu}+m_1 R_\mu^\nu R_\nu^\rho R^\mu _\rho+m_2 R_{\mu\nu}R^{\mu\nu} R+m_3  \,R^3$}

We first deal with the most general Lagrangian involving no derivatives of the Ricci tensor and up to cubic order \cite{Sinha:2010ai}. As there is no derivative of the Ricci in that Lagrangian, the tensor $Z$ is only proportional to $\partial/\partial Riemann$. We have to consider the derivative of the Ricci scalar and Ricci tensor. After symmetrization, we get
\begin{align}
&\frac{\delta R}{\delta R_{abcd}} =\frac12 \left (g^{bd} g^{ac}-g^{ad} g^{bc}\right )\\
&\frac{\delta (R_{\mu\nu} R^{\mu\nu}) }{\delta R_{abcd}} =
\frac 24
\left (g^{bd} R^{ac} -g^{ad} R^{bc}-g^{bc} R^{ad}+g^{ac}R^{bd}\right).
\end{align}
So, 
\begin{align} \nonumber
Z^{abcd}= 
&\left (\frac a2-\frac6\ell^2 b +
\frac{54}{\ell^4} m_3 +\frac{3\nu^2(\nu^2-3)}{\ell^4}m_1+\frac{3\left(3-2 \nu ^2+\nu ^4\right)}{l^4}m_2\right ) 
\left (g^{bd} g^{ac}-g^{ad} g^{bc}\right )
\\
&+ \left (\frac c2 -\frac{3}{2\ell^2}m_2-\frac{3(3+\nu^2)}{4\ell^2}m_1\right )
\left (g^{bd} R^{ac} -g^{ad} R^{bc}-g^{bc} R^{ad}+g^{ac}R^{bd}\right).
\end{align}
Meaning that the constants  $A$ and $B$ are 
\begin{align}
& A=\left (\frac a2-\frac6\ell^2 b +
\frac{54}{\ell^4} m_3 +\frac{3\nu^2(\nu^2-3)}{\ell^4}m_1+\frac{3\left(3-2 \nu ^2+\nu ^4\right)}{l^4}m_2\right ) \\
& B=\left (\frac c2 -\frac{3}{2\ell^2}m_2-\frac{3(3+\nu^2)}{4\ell^2}m_1\right )\,.
\end{align}
For NMG, $a=\frac1 {16\pi}$,$b=\frac{-3}{16\pi 8p }$, $c=\frac1{16\pi p }$ and $m_1=m_2=m_3=0$ with $p= \frac{3-20 \nu ^2}{2 \ell^2}$, so
\begin{equation}
A=\frac{3-5 \nu ^2}{8\pi\left (3 -20  \nu ^2\right )}, \qquad 
B=\frac{\ell ^2}{16\pi\left (3 -20  \nu ^2\right )}\,.
\end{equation}
The charges obtained by \eqref{exactcharges} and \eqref{centralcharges} are consistent with their expressions found by other techniques.

\subsection{Corrections for a Theory with $(\nabla^a R^{bc})^2$}
Take a Lagrangian with the higher curvature term
\be L= \nabla^a R^{bc}\nabla_a R_{bc}.\ee
Following \cite{Azeyanagi:2009wf}, we can compute the $Z$-terms entering the charges:
\bea Z^{abcd}&=&\frac{\partial L}{\partial R_{abcd}}-\nabla_{e_1}\frac{\partial L}{\partial \nabla_{e_1} R_{abcd}}\\
&=&-\frac{1}{2}\left( -g^{ac}\square R^{bd}+g^{bc}\square R^{ad}-g^{bd}\square R^{ac}+g^{ad}\square R^{bc}\right)
\eea
which can be rewritten as \eqref{Z}, using \eqref{eq:useful15}:
\bea
Z^{abcd}&=& \frac{6\nu ^2}{\ell ^4}\left( -g^{ac}g^{bd}+g^{bc}g^{ad}-g^{bd}g^{ac}+g^{ad}g^{bc}\right)\nonumber\\
&&+ \frac{3\nu ^2}{\ell ^2}\left( -g^{ac} R^{bd}+g^{bc} R^{ad}-g^{bd} R^{ac}+g^{ad} R^{bc}\right)
\eea
and
\bea Z^{abcd|e}&=&\frac{\partial L}{\partial \nabla_{e_1} R_{abcd}}\\
&=&\frac{1}{2}\left( g^{ac}\nabla^e R^{bd}-g^{bc}\nabla^e R^{ad}+g^{bd}\nabla^e R^{ac}-g^{ad}\nabla^e R^{bc}\right)\eea
and the relevant corrections to $\FTheta$ and $\textbf{Q}$ are
\bea \FTheta^{(1)}_{a_2a_3}=2\left[\delta g_{ij}\left(R^{j\: i}_{\: b \: c}\nabla^a R^{bc}+R^j_{\: c}(\nabla^a R^{ic}-\nabla^i R^{ac})-R^a_{\: c}(\nabla^j R^{ic}\right)-\delta R_{bc}\nabla^a R^{bc}\right]\e_{aa_2a_3}
\eea
\bea(\Qxi)^{(1)}_{c_3}=-2\xi_k\left[R^b_{\: c}\nabla^a R^{kc}+R^k_{\: c}\nabla^b R^{ac}+R^b_{\: c}\nabla^k R^{ac}\right] \e_{abc_3}\eea
Using the relations \eqref{dJ}, \eqref{RdR}, and $\delta R_{bc}=\frac{\nu^2-3}{\ell^2}\delta g_{bc}$, together with the decomposition of the Riemann tensor, we get the form given by \eqref{AB} in the Appendix:
\bea R^a_{\: c}R^{bc;e}&=&\frac{3\nu}{\ell^5}(\nu^2-1)\Big((\nu^2-3)J^b\e^{ea}_{\:\: s}-2\nu^2 J^a\e^{eb}_{\:\: s}\Big)J^s\\
R^{a\: b}_{\: d \: c}R^{dc;e}&=&\frac{3\nu^3}{\ell^5}(\nu^2-1)\Big(J^b\e^{ea}_{\:\: s}+ J^a\e^{eb}_{\:\: s}\Big)J^s.\eea 
We can then rewrite the corrections as
\bea
 \FTheta^{(1)}_{bc}&=&\frac{6\nu}{\ell^5}(\nu^2-1)\delta g_{ij}\Big(\nu^2 J^i\e^{aj}_{\:\: s} -(2\nu^2-3)J^j\e^{ai}_{\:\: s} -(\nu^2-3)J^i\e^{ja}_{\:\: s} +2\nu^2J^j\e^{ia}_{\:\: s}\Big)J^s\e_{abc}\nonumber\\
 (\Qxi)^{(1)}_{c}&=&
 -\xi_k\frac{6\nu}{\ell^5}(\nu^2-1)\left[(\nu^2-3)\left( J^k\e^{ab}_{\:\: s} +J^a\e^{bk}_{\:\: s} +J^a\e^{kb}_{\:\: s}\right) \right.\nonumber\\
 &&\left. -2\nu^2\left(J^b\e^{ak}_{\:\: s} +J^k\e^{ba}_{\:\: s} +J^b\e^{ka}_{\:\: s} \right)\right]J^s\e_{abc}~.\eea

Since we are going to integrate over $\phi$, and we have to contract the correction to $\Theta$ with $\partial_t$, we need to compute $\Theta^{(1)}_{t\phi}$ and $(\Qxi)^{(1)}_{\phi}$. We also know that $\delta g$ only non-zero components are $\delta g_{rr}$ and $\delta g_{\phi\phi}$. It is then straightforward to show that 
\bea
\FTheta^{(1)}_{t\phi}&=&0\\
(\textbf{Q}_{\partial_t})^{(1)}_{\phi}&=&-72\,r\,\frac{\nu^2}{\ell^6}(\nu^2-1)^2\\
(\textbf{Q}_{\partial_{\phi}})^{(1)}_{\phi}&=&0.
\eea
Even if $(\textbf{Q}_{\partial_t})^{(1)}_{\phi}$ is non-zero, the contribution of this term to the charge vanishes when we take the $\delta$ of it. The same results are recovered using the general method proposed in \cite{Azeyanagi:2009wf}.

\section{Phase Transitions}
To study possible phase transitions between black holes and thermal states, one must first identify the ensemble black holes are dual to. As argued in Sect 2. of \cite{Detournay:2015ysa} or Sect 5.3 of \cite{Detournay:2012pc}, WAdS$_3$ black holes are dual to the ensemble
\be \label{quadens}
   Z = \mbox{Tr} e^{-\beta_+ \tilde{P_0} - \beta_-  \tilde{L_0}} = \mbox{Tr} e^{-\beta Q_T - \beta \Omega  Q_\Phi}
\ee
with 
\be \label{xpxm}
   \tilde{P_0} := Q_{\p_{x^+}} = \frac{P_0^2}{k}, \qquad \tilde{L_0} := Q_{\p_{x^-}} = L_0 - \frac{P_0^2}{k}.
\ee
The coordinates $T$ and $\Phi$ appearing in (\ref{quadens}) are natural coordinates when expressing WAdS$_3$ black holes as a deformation of BTZ black holes\footnote{ The deformation parameter $H$ is given in terms of $\nu$ by $H^2=\frac{3(1-\nu^2)}{2(3+\nu^2)}$, and $\ell_b = 2 \sqrt{\frac1{3 + \nu^2}} \ell$.}::
\be \label{WBTZ}
   ds^2_{WBTZ} = ds^2_{BTZ} - 2 H^2 \xi \otimes \xi
\ee
where $ds^2_{BTZ}$ is the BTZ black hole metric (where we put Newton's constant $G$ to 1):
\be \label{BTZ}
ds^2_{BTZ} = \left(8M-\frac{r^2}{\ell_b^2}\right)  dT^2-\frac{ R^2 dR^2}{8  M R^2-\frac{R^4}{\ell_b^2}-16 J^2 } +8 J\,  dT\,  d\Phi+R^2 d\Phi^2 
\ee
and $\xi$ such that $||\xi||^2 = 1$ is given by
\be \label{xibtz}
\xi^\mu=\frac1{\sqrt8}\sqrt{\frac{ \ell_b}{(M \,  \ell_b -J)} } \left(-\partial_T+\partial_\Phi\right). 
\ee
For the change of coordinates between (\ref{eq:wbhmetric}) and (\ref{WBTZ}), see e.g. Sect 5.4 of \cite{Detournay:2012pc}.


The symmetry argument is still valid in that coordinate system and so the $Z^{abcd}$ field has the same form \eqref{Z}. 
Doing similar computations, the exact charges are given by 
\begin{equation}
Q_T=C\, M\,, \quad
Q_\Phi=C\, J \, . 
\end{equation}
with $C$ a constant depending on the coupling constants, on the deformation parameter $H^2$ and $\ell_b$. 
For example, NMG has  
\begin{equation}
C^{NMG}=\frac{16 \left(1-2 H^2\right)^{3/2}}{17-42 H^2}\,. 
\end{equation}
The thermodynamic potentials are given for any theory by
\begin{equation}
T=  \frac{2 \sqrt{\ell_b^2 M^2-J^2}}{\pi  \ell \sqrt{\ell_b^2\left (\ell_b^2 M^2-J^2\right )}+\ell_b^2M} \,,\quad 
 \Omega= \frac{J}{\sqrt{\ell_b^2\left (\ell_b^2 M^2-J^2\right )}+\ell_b^2M}\,. 
\end{equation}
Integrating the first law, the Gibbs free energy is given by 
\begin{equation}\label{freeenergy}
G=Q_T-T S + \Omega \,Q_\Phi,
\end{equation}
leading to \footnote{WAdS can be obtained from WBTZ by taking $M=-1/8,J=0$. For these values, (\ref{WBTZ}) has an enhanced symmetry and no conical defect/excess.}
\begin{equation}
 G_{WAdS}(T,\Omega)=C\left (-\frac18 \right )\,,\quad  
 G_{WBTZ}(T,\Omega)=C \frac{ -\ell_b^2 \pi ^2 T^2}{2 \left(1-\ell_b^2 \Omega ^2\right)}\,.
 \end{equation}

First, we study the local stability of these phases.  In the grand canonical ensemble, the stability condition is the requirement for a system to have a negative semi-definite Hessian of its free energy $G(T,\Omega)$ . 
This implies that 
\begin{equation}
C>0\,.
\end{equation}
For example in NMG, this condition implies that $H^2<17/42$. 

\begin{figure}\center
                \includegraphics[width=0.5\textwidth]{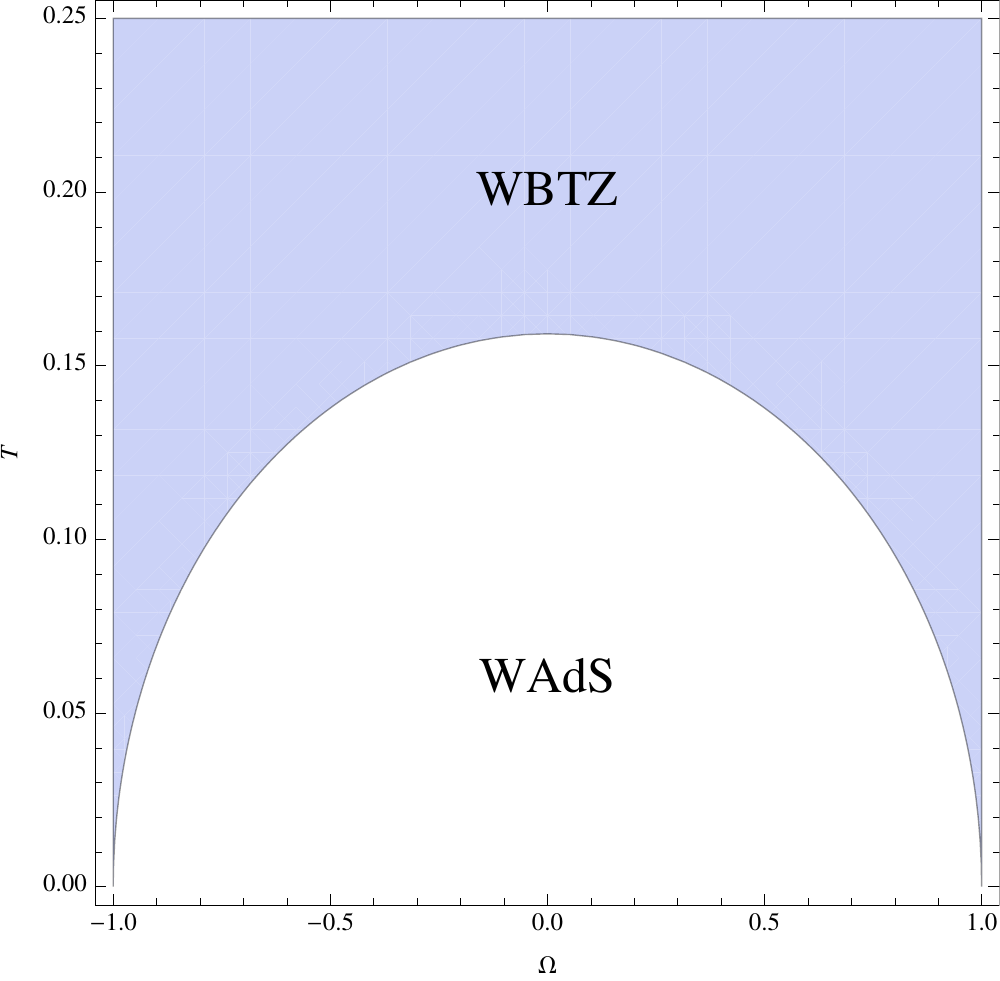}
        \caption{WBTZ-WAdS phase diagram. WBTZ dominates in the purple section and WAdS in the white.}\label{fig:BTZ-AdS-TMG}
        \label{phasediagr-warped}
\end{figure}

Secondly, we consider the global stability.
In the classical limit, the dominant phase is the most probable, i.e. the one who dominates the partition function among the saddle points. 
Here the two known phases are the black hole and the thermal vacuum. We compare their free energies through their difference
\begin{equation}
\Delta G =C\left(-\frac{1}{8}+\frac{\ell_b^2\text{  }\pi ^2 T^2}{ 2\left(1-\ell_b^2 \Omega ^2\right)}\right)\,. 
\end{equation}
If $\Delta G<0$, WAdS dominates and for $\Delta G>0$, WBTZ dominates.
The constant $C$ factorizes out \footnote{We take $C>0$ because it is meaningless to consider the global stability of a phase that is not locally stable.}. 
It implies that the phase diagram does not depend on which theory we look at and moreover, as thermodynamic potentials also do not depend on the deformation parameter, it is the same phase diagram as for BTZ black holes.
The phase diagram is represented (for $\ell=1$) in figure \ref{phasediagr-warped}. 
Our results differ from those of \cite{Ghodrati:2016ggy} in which the authors dealt with the NMG case, but considered a different ensemble.

\section*{Acknowledgements}
We would like to thank Tatsuo Azeyanagi, Gaston Giribet and Aninda Sinha for useful discussions and correspondence. G. N. was supported by an NSERC Discovery Grant. L.-A. D. is supported by a WBI.World fellowship of the Fondation Wallonie-Bruxelles International. S.D. and C.Z. are supported in part by the ARC grant ``Holography, Gauge Theories and Quantum Gravity -- Building models of quantum black holes", by FNRS-Belgium (convention IISN 4.4503.15) and benefited from the support of the Solvay Family. S.D. is a Research Associate of the Fonds de la Recherche Scientifique F.R.S.-FNRS (Belgium). C.Z. is a research fellow of ``Fonds pour la Formation et la Recherche dans l'Industrie et dans l'Agriculture"- FRIA Belgium.
This work was partially supported by IISN - Belgium (convention 4.4504.15).
\appendix

\section{On-shell Conditions for Theories without Derivatives of Riemanns}
\label{app:on-shell}
In this section, we shall study the consequences of the equations of motion for the most general theory without derivatives of Riemanns. Moreover, we will show that on-shell conditions imply Eq.~(\ref{relationAB}).
Due to the fact that we are in three dimensions, the most general action (without derivatives of Riemanns) is of the form
\be
I= \int d^3 x \sqrt{-g} f(R_{\mu\nu}, g^{\mu\nu}) \,.
\ee 
The equation of motion is easily derived (see \cite{Saida:1999ei} for example) to be
\be\label{eq:eomappA}
2\frac{\partial f}{\partial g^{\mu\nu}} -  f g_{\mu\nu}
=
\nabla^\alpha \nabla_\nu P_{\alpha\mu}
+\nabla^\alpha \nabla_\mu P_{\alpha\nu}
-\square P_{\mu\nu}
-g_{\mu\nu} \nabla^\beta \nabla^\alpha P_{\alpha\beta}
\ee where
\be
P_{\mu\nu} = g_{\mu\alpha} g_{\nu\beta} \frac{\partial f}{\partial R_{\alpha\beta}}.
\ee
The object of interest $Z^{\mu\nu\alpha\beta}$ in three dimensions is
\bea
Z^{\mu\nu\alpha\beta}&\equiv& \frac{\partial L}{\partial R_{\mu\nu\alpha\beta}}
=
\frac{\partial R_{\gamma\delta}}{\partial R_{\mu\nu\alpha\beta}} \frac{\partial L}{\partial R_{\gamma\delta}}\nonumber\\
&=&
\frac{1}{4}
\left[
g^{\mu\alpha}   \delta_{\delta}{}^{ \beta} \delta_{\gamma}{}^{ \nu}
-g^{\alpha\nu} \delta_{\gamma}{}^{ \mu}  \delta_{\delta}{}^{ \beta}
+g^{\beta\nu} \delta_{\gamma}{}^{ \mu} \delta_{\delta}{}^{ \alpha}
-g^{\beta\mu} \delta_{\gamma }{}^{\nu}  \delta_{\delta}{}^{ \alpha}
\right]\frac{\partial L}{\partial R_{\gamma\delta}}\,
\eea  which is related to $P^{\mu\nu}$ by
\be\label{eq:ZappA}
 Z^{\mu\nu\alpha\beta}=
\frac{1}{2}
\left[
g^{\mu\alpha}   g^{ \beta \nu} 
-g^{\alpha\nu} g^{\beta \mu}  
\right]+
\frac{1}{4}
\left[
g^{\mu\alpha}   P^{\beta\nu}
-g^{\nu\alpha} P^{\beta \mu}
+g^{\beta\nu} P^{\alpha \mu}
-g^{\beta\mu} P^{\alpha\nu}
\right]\,.
\ee

For later purposes, it is useful to note that for a locally WAdS$_3$ spacetime, due to $SL(2,\mathbb{R})\times U(1)$ symmetry, we have that for $q\ge 1$,
\be \label{eq:Rmnq}
(R_{\alpha\beta}^{q})^{\mu \nu}
\equiv R_{\alpha \beta_1} R^{\beta_1}{}_{\beta_2} R^{\beta_2}{}_{\beta_3}\ldots
R^{\beta_{q-1}}{}_\beta
={A}_{q} g^{\mu\nu} + {B}_{q} R^{\mu\nu},\quad
\ee with $A_q$ and $B_q$ constants which are dependent on $\ell$ and $\nu$. 
For example, $ A_{1}=0,\quad B_1=1,\quad A_2=2 \nu ^2 \left(\nu ^2-3\right) /\ell^4,\quad B_2=-(3+\nu ^2)/\ell^2$.
By definition, the $A_q$ and $B_q$ satisfy the following recursion relations
\bea
A_q&=&A_2 B_{q-1}  = \frac{2\nu^2 (\nu^2-3)}{\ell^4} B_{q-1} \\
\label{eq:Bquseful}
B_q&=&A_{q-1}+ B_2 B_{q-1}   = A_{q-1}-\frac{3+\nu ^2}{\ell^2} B_{q-1}\,.
\eea
As a notational convention, we will denote $\tr(R_{\alpha\beta}^q)\equiv (R_{\alpha\beta}^{q})^{\mu}{}_\mu$.
Moreover, it is also useful to note that
\bea\label{eq:useful15}
\nabla^\alpha \nabla_\mu R_{\alpha\nu}=\nabla^\alpha \nabla_\nu R_{\alpha\mu}&=& -\frac{6 \nu ^2}{\ell^4}g_{\mu\nu} 
-\frac{3 \nu ^2}{\ell^2}R_{\mu\nu} \nonumber\\
\square R_{\mu\nu}&=&\frac{12 \nu ^2}{\ell^4}g_{\mu\nu} +\frac{6 \nu ^2}{\ell^2} R_{\mu\nu}\,,
\eea and so
\bea\label{eq:useful16}
\nabla^\alpha \nabla_\mu R_{\alpha\nu}+\nabla^\alpha \nabla_\nu R_{\alpha\mu}-\square R_{\mu\nu}&=&-\frac{24 \nu ^2}{\ell^4}g_{\mu\nu}-\frac{12 \nu ^2}{\ell^2} R_{\mu\nu}
\eea while $\nabla_\alpha R^{\alpha\beta}=0$ using the contracted Bianchi identity and the fact that $R$ is a constant. 

For illustrative purposes, let us first consider the simple case where for some fixed $k\ge 2$,
\be
f=f_k\equiv c_k R^k+b_k  ~\tr(R_{\mu\nu}^k)\,.
\ee In this case,
\be\label{eq:Peg1}
P_{\mu\nu} = k\left[ c_k  g_{\mu\nu}  R^{k-1} +b_k  (R_{\alpha\beta}^{k-1})_{\mu\nu}\right]
\ee while
\bea
Z^{\mu\nu\alpha\beta}
&=&
\frac{k}{2}\left[
c_k  R^{k-1} +b_k  A_{k-1}
\right]
\left[
g^{\mu\alpha}   g^{ \beta \nu} 
-g^{\alpha\nu} g^{\beta \mu}  
\right]
\nonumber\\
&+&
\frac{k}{4}b_k  B_{k-1}
\left[
g^{\mu\alpha}   R^{\beta\nu}
-g^{\nu\alpha}   R^{\beta\mu}
+g^{\beta\nu}   R^{\alpha\mu}
-g^{\beta\mu}  R^{\alpha\nu}
\right] 
\eea where we have used Eq.~(\ref{eq:Rmnq}).
On the other hand, the equation of motion Eq.~(\ref{eq:eomappA}) in this case reads
\bea
&&
k\left[ c_k  R^{k-1}R_{\mu\nu} 
+ b_k(R_{\alpha\beta}^k)_{\mu\nu}\right]
- \half \left[ c_k R^k 
+ b_k\tr(R_{\alpha\beta}^k) \right]g_{\mu\nu}
\nonumber\\
&=&
\half k b_k\left\{\nabla^\alpha \nabla_\nu[ (R_{pq}^{k-1})_{\alpha\mu}]
+\nabla^\alpha \nabla_\mu[  (R_{pq}^{k-1})_{\alpha\nu}]
-\square [ (R_{pq}^{k-1})_{\mu\nu}]
-g_{\mu\nu} \nabla^\beta \nabla^\alpha [ (R_{pq}^{k-1})_{\alpha\beta}] \right\}\nonumber\\
&&+k c_k \left[  \nabla_\mu \nabla_\nu R^{k-1} 
- g_{\mu\nu} \square  R^{k-1} \right]
\eea which upon using Eq.~(\ref{eq:Rmnq}) and Eq.~(\ref{eq:useful15})-(\ref{eq:useful16}) yields
\bea\label{eq:eom11}
0&=&
\left[k c_k  R^{k-1}+k b_k B_k
+ k b_k B_{k-1}
\frac{6 \nu ^2}{\ell^2} 
\right]R_{\mu\nu} \nonumber\\
&&
- \half \left[
 c_k R^k 
-2k b_k A_k 
+ b_k\tr(R_{\alpha\beta}^k) 
-k b_k B_{k-1}
\frac{24 \nu ^2}{\ell^4}
\right]g_{\mu\nu} \,.
\eea
By explicitly plugging in the metric and $R_{\mu\nu}$ for a locally WAdS$_3$ metric, these equations in turn become two decouple equations
\bea\label{eq:eom1}
0&=&
 c_k  R^{k-1}+ b_k B_k
+  b_k B_{k-1}
\frac{6 \nu ^2}{\ell^2} \,,
\\
 0&=&c_k R^k 
-2k b_k A_k 
+ b_k\tr(R_{\alpha\beta}^k) 
-k b_k B_{k-1}
\frac{24 \nu ^2}{\ell^4}\,.
\eea

Let us now recall from Eq.~(\ref{eq:ZappA}), Eq.~(\ref{eq:Rmnq}) and Eq.~(\ref{eq:Peg1}) that in this case we have
\bea
Z^{\mu\nu\alpha\beta}
&=&
\frac{k}{2}\left[
c_k  R^{k-1} +b_k  A_{k-1}
\right]
\left[
g^{\mu\alpha}   g^{ \beta \nu} 
-g^{\alpha\nu} g^{\beta \mu}  
\right]
\nonumber\\
&+&
\frac{k}{4}b_k   B_{k-1}
\left[
g^{\mu\alpha}   R^{\beta\nu}
-g^{\nu\alpha}   R^{\beta\mu}
+g^{\beta\nu}   R^{\alpha\mu}
-g^{\beta\mu}  R^{\alpha\nu}
\right] \nonumber\\
&\equiv &
A
\left[
g^{\mu\alpha}   g^{ \beta \nu} 
-g^{\alpha\nu} g^{\beta \mu}  
\right]
+B\left[
g^{\mu\alpha}   R^{\beta\nu}
-g^{\nu\alpha}   R^{\beta\mu}
+g^{\beta\nu}   R^{\alpha\mu}
-g^{\beta\mu}  R^{\alpha\nu}
\right] \,,
\eea
where 
\be
A\equiv \frac{k}{2}\left[
c_k  R^{k-1} +b_k  A_{k-1}
\right],\quad B \equiv\frac{k}{4}b_k   B_{k-1}\,.
\ee  Their ratio is
\be
\frac{B}{A}
=\frac{1}{2} \frac{b_k   B_{k-1}}{
c_k  R^{k-1} +b_k  A_{k-1}
}
=\frac{1}{2} \frac{ 1  }{
- 6\nu^2/\ell^2
+ ( A_{k-1}-B_k)/B_{k-1}
}
\ee where we have used one of the equations of motion Eq.~(\ref{eq:eom1}).
Using Eq.~(\ref{eq:Bquseful}), we obtain
\be
\frac{B}{A}
=-\frac{\ell^2}{
2(-3+5\nu^2)
}
\ee which is Eq.~(\ref{relationAB}) as required by finiteness of charges.

Moreover, one can straightforwardly generalize the same computations to a more general $f=f_{k;q_1,\ldots,q_n}\equiv c(k,q_1,\ldots q_n) R^k \times\tr[(R_{\mu_1\nu_1})^{q_1}] \times\tr[(R_{\mu_2\nu_2})^{q_2}]\times\ldots \times\tr[(R_{\mu_n\nu_n})^{q_n}]$ or even the most general action 
\be
f=\sum_{k,n}\sum_{q_1,\ldots,q_n} f_{k;q_1\ldots q_n}.
\ee The upshot is that eventually similar arguments as above follow through and imply
Eq.~(\ref{relationAB}) as desired.

\section{Computations of  
$\Theta^{(s)}$ and $Q_\xi^{(s)}$ (for $s\ge1$) }
\label{sec:appTheta}
In this appendix, we explicit the computation of the corrections to the charges.
First, all of terms appearing in $\Theta^{(s)}$ and $(\Qxi^{(s)})$ are of the following form\footnote{There is a term in the expression of $\Theta^{(s)}$ that looks like $Z\delta R$ which seems like it cannot be manifestly written in the form in Eq.~(\ref{eq:structure1}). However, in Sec.~(\ref{sec:lastterm}), we show that with a bit of work, even this term can be put into the form in Eq.~(\ref{eq:structure1}).}
 \bea\label{eq:structure1}
(\Qxi)^{(s)}_{c_3\ldots c_n}& =&\xi_k A^{kab}\epsilon_{ab c_3 \ldots c_n}\nonumber\\
\FTheta^{(s)}_{a_2\ldots a_n}& =&( \delta g_{ij}) B^{ija}\epsilon_{a a_2\ldots a_n}\,,
\eea where $B^{ija}$ is a tensor symmetric in the $(i,j)$ indices while $A^{kab}$ is antisymmetric in $(a,b)$. Both $A$ and $B$ are covariant tensors constructed out of the metric. By $SL(2,\mathbb{R}) \times U(1)$ symmetry, in the vielbein-analysis, we know that by boost-invariance, there are four independent non-zero components for $B^{ija}$ while there are three for $A^{kab}$.
This means that we can decompose them in the following way
\bea \label{AB}
A^{abk}&=&\left ( \left (a_1 g^{ak}+a_2 \epsilon^{akc}J_c\right )J^b -(a\leftrightarrow b)\right )
+a_3 \epsilon^{abp} J_p J^k
\nonumber\\
B^{ija}&=& b_1 g^{ij}J^a+\left( \left(b_2J^i g^{ja}+b_3J^i \epsilon^{jak}  J_k \right)  +(i\leftrightarrow j)\right) 
+b_4 J^i J^j J^a  \,
\eea  
where $a_1,a_2,a_3$ are constants depending only on $\nu$ and $\ell$.

Let us first focus on the expression for $\FTheta^{(s)}$. Using the fact that the only non-zero components of $\delta g_{ij}$ are the $\delta g_{rr}$ and $\delta g_{\phi\phi}$ components and that $g^{ij}\delta g_{ij}=0$, we get
\be
\label{eq:structure2}
\FTheta^{(s)}_{a_2\ldots a_n} =
( \delta g_{ij})\left (\left( \left(b_2J^i g^{ja}+b_3J^i \epsilon^{jak}  J_k \right)  +(i\leftrightarrow j)\right)
+b_4 J^i J^j J^a \right ) \epsilon_{a a_2\ldots a_n} \,.
\ee  
Moreover $J^{\mu}\partial_\mu = \partial_t$ implies that the first four terms in the bracket of Eq.~(\ref{eq:structure2}) (proportional to either $\delta g_{ij} J^i$ or $\delta g_{ij} J^j$) do not contribute to $\FTheta^{(s)}$.
 Furthermore, since we are interested in the $a=r$ component (to compute charges at $r$=constant surface), the last term does not contribute either. Hence, we have established that in the computation of $\delta\ffL0$ and $\delta\ffP0$, $\FTheta^{(s)}$ vanishes.

Next, consider the expression for $(\Qxi)^{(s)}$ 
\begin{equation}\label{eq:structure3}
(\Qxi)^{(s)}_{c_3\ldots c_n}=\xi_k \left (a_3 \epsilon^{abp} J_p J^k+ \left(2a_1 g^{k[a}+2a_2 \epsilon^{pk[a} J_p\right)J^{b]}\right )\epsilon_{ab c_3 \ldots c_n}\,.
\end{equation}
 For $\delta P_0$, we consider $\xi^\mu\partial_\mu = \partial_t=J^\mu\partial_\mu$. We see that the second and third term in the square bracket in Eq.~(\ref{eq:structure3}) vanish since they are proportional to $\epsilon_{ab} J^a J^b=0$. Thus we are left to evaluate
\be \label{Qdt}
\int_{\infty }^{}(\Qxi)^{(s)}
=\int_\infty \xi_k \left (2a_2 \epsilon^{pk[a} J_pJ^{b]}\right)\epsilon_{ab c}dx^c 
= \frac{8\pi \nu \, a_3}{\ell} r  \,,
\ee where we have used $J^\mu J_\mu=1$ and $J_\mu dx^\mu=dt-(2r\nu/\ell)d\phi$ while setting the $(a,b)=(t,r)$. 
We should still take the $\delta$ of that expression, so it gives us zero. 

Finally, we consider the expression for $(\Qxi)^{(s)}$ for $\xi^\mu\partial_\mu = \partial_\phi$ in the computation of $\delta \ffL0$. Direct computation shows
\be
\int_{\infty}(\Qxi)^{(s)}=
%
\left (\frac{16\,\pi \,a_3 \,\nu ^2+4 \,\pi \,a_2\left(3+\nu ^2\right)}{\ell^2}\right ) r^2
-\left (48\, a_2 \,m \right )r
+\frac{16\, a_2\, \ell }{\nu }j\,.
\ee 
%
%
%
The corrections to the central charges are proportional to the Lie derivative of the Noether charge. Explicit computations gives us the following
\begin{equation}
c^{(s)}=\frac{192 \left(3 \,a_2 \,\pi +a_2 \,\pi  \,\nu ^2+4\, a_3 \,\pi \, \nu ^2\right) r^2}{l^2}-1152 \,a_2 \,m \,\pi \, r\,,\qquad 
k^{(s)}= -16\,\pi\, a_3\,.
\end{equation}

\subsection{Putting the $Z\delta R$ term in the form of Eq.~(\ref{eq:structure1})}
\label{sec:lastterm}
Consider the following object
\be
T^a\equiv  Z^{kbcd|e_1\ldots e_{s-1} a} \delta \left[\mathbb{R}_{kbcd;e_1\ldots e_{s-1}}\right]
\ee which is one of the terms appearing in $\FTheta^{(s)}$ that does not look like of the form Eq.~(\ref{eq:structure1}).
Since every covariant tensor built out of a $SL(2,\mathbb{R})$ metric can be written in terms of polynomials of $\epsilon, g_{\mu\nu}$ and $J_\rho$ appropriately. Then
\be
\mathbb{R}_{kbcd;e_1\ldots e_{s-1}}=\sum_{p} c_p(\nu,\ell) t^{(p)}_{kbcd;e_1 \ldots e_{s-1}}
\ee where $t$ is some basis tensor built out of polynomials of $\epsilon,g$ and $J$. Crucially the coefficient $c_p$ only depends on $(\nu,l)$ and not on the black hole or quotienting parameters. Note that
\be
\delta J_\mu=\delta \epsilon_{abc}=0,
\ee since $\sqrt{-g}=1$ and $J_\mu dx^\mu = dt-(2\nu r/\ell) d\phi$  does not contain black hole's parameters. Therefore, we have
\be
\delta \mathbb{R}_{kbcd;e_1\ldots e_{s-1}}=\sum_{p} c_p(\nu,\ell) \delta t^{(p)}_{kbcd;e_1 \ldots e_{s-1}}=\sum_p c_p(\nu,\ell) \frac{\delta t^{(p)}_{kbcd;e_1 \ldots e_{s-1}}}{\delta g_{ij}} \delta g_{ij},
\ee since variations hit the $t$-tensor through $g_{\mu\nu}$ and that $t$ is a {\it polynomial} of $g$. It is important to note that no covariant derivatives of $g_{\mu\nu}$ appear in this $t$.
Therefore, we obtain
\be
T^a\equiv  \left[Z^{kbcd|e_1\ldots e_{s-1} a}
\sum_p c_p
\frac{\delta t^{(p)}_{kbcd;e_1 \ldots e_{s-1}}}{\delta g_{ij}} \right]\delta g_{ij}
\equiv A^{ija} \delta g_{ij},
\ee where $A^{ija}=A^{jia}$.

\bibliographystyle{utphys}
\bibliography{3dwarpedBHbibV2}

\end{document}